\documentclass[preprint,3p]{elsarticle}

\usepackage[utf8]{inputenc}
\usepackage{amsmath,amssymb,amsfonts}
\usepackage{graphicx}
\usepackage{bbm}
\usepackage{tabularx}
\usepackage{booktabs}
\usepackage{makecell}
\usepackage{xcolor}
\usepackage[detect-weight=true]{siunitx}

\DeclareMathOperator{\sgn}{sgn}
\DeclareMathOperator*{\argmin}{arg\,min}
\newcommand{\RR}{\mathbb{R}}

\begin{document}

\begin{frontmatter}

    \title{The WQN algorithm to adaptively correct artifacts in the EEG signal\tnoteref{funding}}

    \author[1]{Matteo Dora}
    \ead{matteo.dora@ens.psl.eu}

    \author[2]{Stéphane Jaffard}
    \ead{jaffard@u-pec.fr}

    \author[1]{David Holcman \corref{corrauth}}
    \ead{david.holcman@ens.psl.eu}
    \cortext[corrauth]{Corresponding author: david.holcman@ens.psl.eu.}

    \affiliation[1]{%
        organization={IBENS UMR8197, \'Ecole Normale Sup\'erieure -- PSL},
        postcode={75005},
        city={Paris},
        country={France}
    }
    \affiliation[2]{%
        organization={LAMA UMR8050, Univ Paris Est Créteil, CNRS},
        postcode={F-94010},
        city={Créteil},
        country={France}
    }

    \tnotetext[funding]{Funding: D.H.'s research is supported by grants ANR NEUC-0001, PSL and CNRS pre-maturation, and by the European Research Council (ERC) under the European Union's Horizon 2020 research and innovation program (grant agreement Nº~882673).}

    \begin{keyword}
        wavelet quantile normalization \sep wavelet transform \sep transport \sep wavelet thresholding \sep EEG \sep artifact removal
    \end{keyword}

    \begin{abstract}
        Wavelet quantile normalization (WQN) is a nonparametric algorithm designed to efficiently remove transient artifacts from single-channel EEG in real-time clinical monitoring. Today, EEG monitoring machines suspend their output when artifacts in the signal are detected. Removing unpredictable EEG artifacts would thus allow to improve the continuity of the monitoring. We analyze the WQN algorithm which consists in transporting wavelet coefficient distributions of an artifacted epoch into a reference, uncontaminated signal distribution. We show that the algorithm regularizes the signal. To confirm that the algorithm is well suited, we study the empirical distributions of the EEG and the artifacts wavelet coefficients. We compare the WQN algorithm to the classical wavelet thresholding methods and study their effect on the distribution of the wavelet coefficients. We show that the WQN algorithm preserves the distribution while the thresholding methods can cause alterations. Finally, we show how the spectrogram computed from an EEG signal can be cleaned using the WQN algorithm.
    \end{abstract}
\end{frontmatter}


\section{Introduction}

\noindent Artifact removal constitutes a major challenge in the processing of real-time brain time series such as electroencephalogram (EEG): how to suppress transient large amplitude artifacts without irreparably altering the statistical properties of the underlying signal? This situation is in contrast with traditional denoising approaches, which are designed to remove small additive noise contaminating the signal. For this latter class of problems, wavelet-based methods~\cite{Mey90I,jaffard2001wavelets,donoho1995wavelet,averbuch1996image,BCM} have been shown to be very effective at identifying multiscale time-frequency patterns, allowing to separate signal from noise. The wavelet transform (WT) is now commonly used in many applications ranging from images~\cite{averbuch1996image}, audio~\cite{OmerTorr}, to physiological signals~\cite{unser1996review,le2007analysis,worrell2012recording,Lina,CiuCiuetal}.

Thresholding of small wavelet coefficients have been classically used to remove small amplitude noise with low computational cost. However, this approach requires to estimate the statistical properties of the signals in order to obtain a clear separation of the signal from noise~\cite{uriguen2015eeg}. Various methods have been proposed to optimally define threshold values based on these statistical properties~\cite{donoho1995wavelet,donoho1998minimax}, focusing on signals contaminated by small additive white noise. The wavelet thresholding approach has been adapted to remove artifacts from EEG~\cite{krishnaveni2006removal,inuso2007waveletica,chavez2018surrogate} but with opposite assumptions: the physiological signal is represented by a small-amplitude, stationary process which is corrupted by large and isolated artifacts. Contrary to traditional applications of wavelet thresholding, artifact removal is obtained by suppressing a small number of \emph{large} wavelet coefficients.
In this context, the purpose of denoising is different: it is no more to restore the point by point values of the initial signal, but rather its texture, i.e.\ its initial statistical properties.

As an alternative to wavelet thresholding, we recently introduced the wavelet quantile normalization (WQN) algorithm~\cite{dora2022adaptive} to remove artifacts from single-channel EEG in real-time clinical monitoring. The WQN method consists in transporting the distributions of wavelet coefficients belonging to an artifacted signal fragment onto a reference distribution extracted by an adjacent uncontaminated fragment. In this letter, we compare the thresholding approach and the WQN algorithm when used to remove large transient artifact from EEG time series, highlighting the properties of the latter algorithm in terms of preservation of wavelet coefficients statistics and regularization. After having briefly recalled the WQN algorithm (in Section~\ref{sec:wqn_algorithm}), in Section~\ref{sec:eeg_statistics} we present wavelet coefficients statistics for two classes of EEG artifacts and compare them to those of physiological EEG.\@ In Section~\ref{sec:wt_comparison}, we compare the WQN to wavelet thresholding methods, showing how the latter can cause significant alteration of the coefficient statistics when they are employed backwards by suppressing large coefficients. In Section~\ref{sec:regularity} we study the regularity properties of the WQN algorithm, showing that it cannot introduce unwanted singularities in the signal. Finally, in Section~\ref{sec:application}, we discuss the application of WQN to remove artifact from EEG during anesthesia in the surgery room. Uninterrupted monitoring of these statistics can provide a continuous guidance for anesthetists in taking the relevant decisions.


\section{Description of the WQN algorithm}\label{sec:wqn_algorithm}

\begin{figure}
    \includegraphics[width=\linewidth]{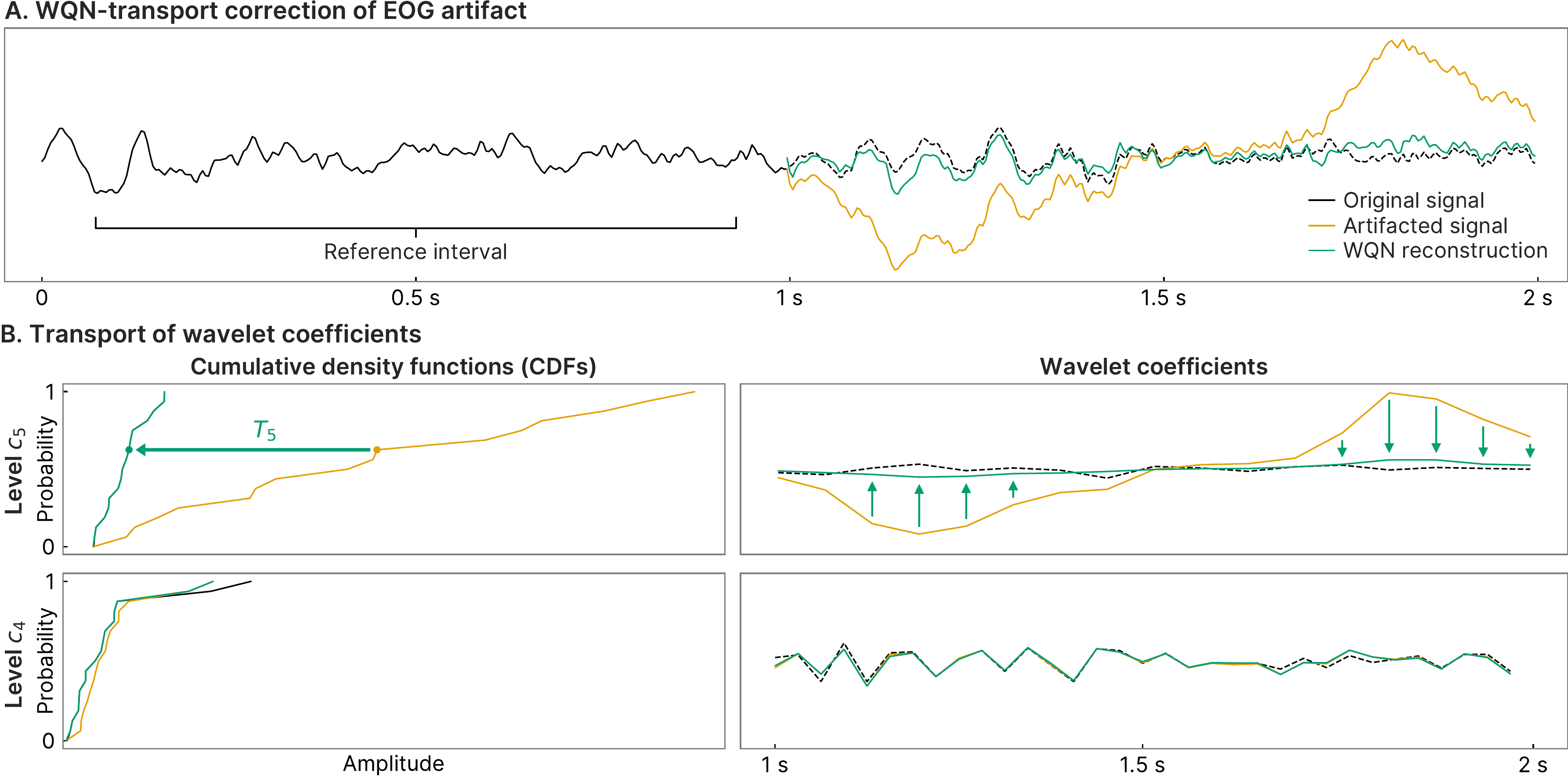}
    \caption{WQN correction of a simulated eye-movement artifact. \textbf{(A)}~Reconstruction of the original signal based on WQN (green) from the artifacted signal (orange), compared to the original signal (black). \textbf{(B)}~Effect of the coefficient normalization at scales $c_4$ and $c_5$, showing the empirical CDF transport (left) and the attenuation of the wavelet coefficients (right) as described by Eq.~\ref{eq:coeff_normalization}. The algorithm adaptively attenuates coefficients at scale $c_5$ (where most of the artifact power is concentrated) leaving coefficients $c_4$ mostly unmodified (as this scale is almost unaffected by the artifact), demonstrating the relevance of a scale-by-scale normalization approach.}%
    \label{fig:exampleWQN}
\end{figure}

\noindent The wavelet quantile normalization (WQN) algorithm~\cite{dora2022adaptive} was designed to eliminate transient artifacts~\cite{tatum2011artifact,schomer2012niedermeyer} from single-channel EEG signals. The method works by normalizing the wavelet coefficients of artifacted intervals using reference statistics extracted from an uncontaminated signal fragment, typically adjacent to the artifacted interval (Fig.~\ref{fig:exampleWQN}A). The artifacted and reference intervals are decomposed with an M-level discrete wavelet transform (DWT) defined as~\cite{daubechies1992ten}:
\begin{equation}\label{eq:dwt}
    x(t) = \sum_{m = 1}^{M} \sum_n c_{m,n} \, \psi_{m,n}(t) + \sum_n c_{M + 1,n} \, \phi_{M,n}(t),
\end{equation}
where $\psi_{m, n}\left(t\right) = 2^{-m / 2} \, \psi\left(2^{-m}\, t - n\right)$, $\phi_{m, n}\left(t\right) = 2^{-m / 2} \, \phi\left(2^{-m}\, t - n\right)$, represent wavelet and scaling functions, and $c_{M+1,n} = \left< x, \phi_{M,n} \right>$, $c_{m\leq M,n} = \left< x, \psi_{m,n} \right>$ are the approximation and detail coefficients.

For each level $m$, the coefficients $c^{\text{(art)}}_{m}$, $c^{\text{(ref)}}_{m}$ are used to compute the empirical cumulative density functions (CDF) $F^{\text{(ref)}}_{m}$, $F^{\text{(art)}}_{m}$ of the coefficients amplitude for artifacted and reference intervals respectively:
\begin{align}
    F_{m}(x) = \frac{1}{N_m} \sum_{n = 1}^{N_m}  \mathbbm{1}_{\left| c_{m,n} \right| <\,x},
\end{align}
where $N_m$ indicates the number of coefficients and $\mathbbm{1}_{\left| c \right| <\,x}$ is the indicator function which takes value 1 if $\left| c \right| <\,x$ and 0 otherwise. The wavelet coefficients $c^{\text{art}}_m$ are transformed such that the distribution of their amplitude matches that of the reference interval, via the mapping $T_m$ defined as
\begin{equation}\label{eq:cdf_map}
    T_m(x) = F^{\text{(ref)}\, -1}_{m} \left( F^{\text{(art)}}_{m}(x) \right),
\end{equation}
where $F^{\text{(ref)} \, -1}_{m}$ indicates the generalized inverse of $F^{\text{(ref)}}_{m}$ (in the sense of completed graphs for discontinuous increasing functions, see Fig.~\ref{fig:exampleWQN}). This mapping transports the amplitude distribution of $c^{\text{(art)}}_m$ to the distribution of the reference interval signal.

Finally, the normalization function
\begin{align}\label{eq:coeff_normalization}
     & \lambda_{m}(c) = \sgn(c) \cdot \min\left\{\left| c \right|,\,
    T_m \left( \left| c \right| \right)
    \right\},
\end{align}
maps a coefficient $c$ from $c^{\text{(art)}}_m$ to its possibly attenuated value. Equation~\ref{eq:coeff_normalization} ensures that the norm of wavelet coefficients is never increased, a key requirement to guarantee the regularity of the algorithm as we will see in Sec.~\ref{sec:regularity}.
The corrected coefficients are defined by
\begin{equation}\label{eq:corrected_coeffs}
    c^{\text{(corr)}}_{m,n} =  \lambda_m \left( c^{\text{(art)}}_{m,n} \right), \quad m = 1, \dots, M + 1.
\end{equation}
The corrected version of the artifacted segment is obtained by inverting the DWT using the corrected coefficients
\begin{equation}\label{eq:dwt_reconstruction}
    x^{\text{(corr)}}(t) = \sum_{m = 1}^{M} \sum_n c^{\text{(corr)}}_{m,n} \, \psi_{m,n}(t) + \sum_n c^{\text{(corr)}}_{M+1,n} \, \phi_{M,n}(t).
\end{equation}

To illustrate how the WQN algorithm performs, we added an eye-movement artifact (EOG, electrooculogram) to an EEG signal (Fig.~\ref{fig:exampleWQN}A), showing how the algorithm transports wavelet coefficients from the artifacted to the reference distribution (Fig.~\ref{fig:exampleWQN}B). The single-level coefficient mapping (Fig.~\ref{fig:exampleWQN}B) illustrates how the algorithm adapts to different scales.

\section{Statistics of EEG and artifact signals}\label{sec:eeg_statistics}

\begin{figure*}
    \includegraphics[width=\textwidth]{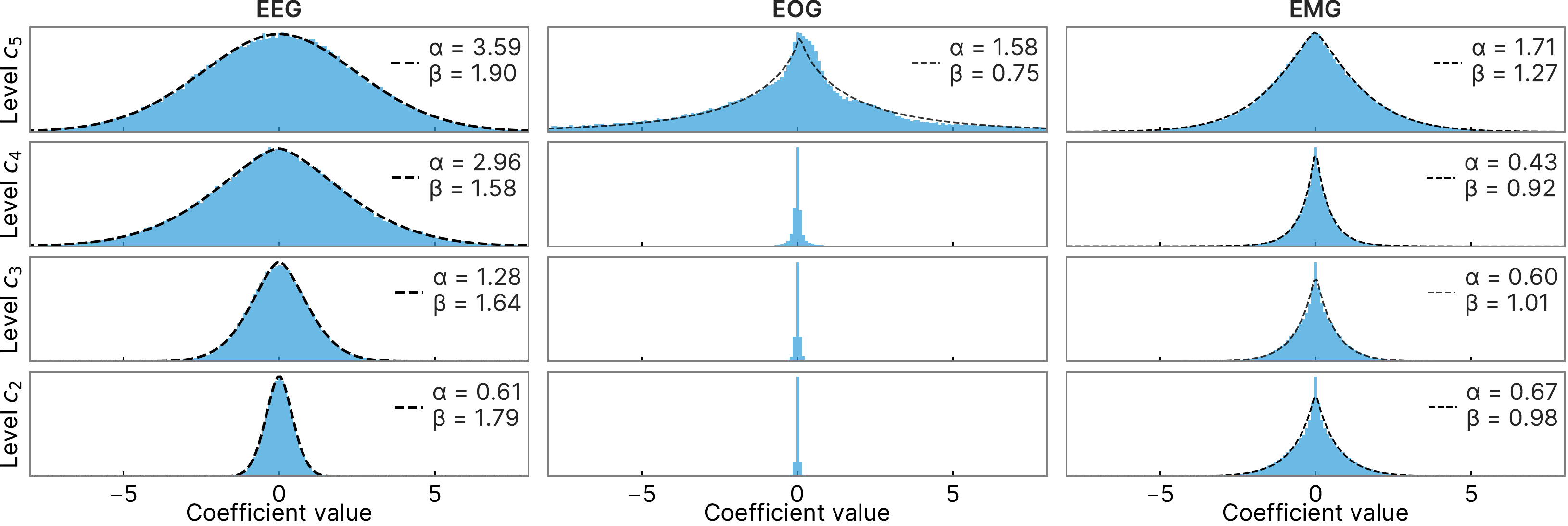}
    \caption{
        Distribution of the wavelet coefficients for EEG, electrooculogram (EOG) and electromyogram (EMG). The wavelet distribution for the EEG signal can be well approximated by a centered generalized Gaussian distribution (dashed line), fitted by maximum-likelihood~\cite{varanasi1989parametric}.
        The statistics were computed on EEG, EOG, and EMG recordings taken from the Denoise-Net dataset~\cite{zhang2020eegdenoisenet}. Before computing the wavelet transform, each signal was normalized to have standard deviation equal to 1.
    }%
    \label{fig:distributions}
\end{figure*}

\noindent We further studied the distributions of the wavelet coefficients of the EEG signal and compared them to artifacts generated by eye movement (EOG) and muscular activity (EMG). We decomposed 2-second signals from the Denoise-Net dataset~\cite{zhang2020eegdenoisenet} using a symlet wavelet with 5 vanishing moments to obtain the coefficient distributions for each scale (Fig. \ref{fig:distributions}). Interestingly, the wavelet coefficients of the unperturbed EEG can be well described by generalized Gaussian distributions ($f(x) = \frac{\beta}{2 \alpha \Gamma (1/\beta)} \exp\left[ - {\left( |x| / \alpha \right)}^\beta \right]$) with shape parameter $\beta$ between 1.5 and 2 (i.e. a standard Gaussian). This is in contrast with the coefficient distribution of the artifacted epochs. In EOG, non-zero coefficients are concentrated in the low frequency scale (level $c_5$); their distribution is peaked around small coefficients and shows a clear deviation from a generalized Gaussian due to skewness (Fig. \ref{fig:distributions}, second column). In EMG, non-zero coefficients span several scales (Fig. \ref{fig:distributions}, third column), similarly to the EEG, but they have a peaked distribution (with $\beta$ close to 1, i.e. Laplace distribution).
In summary, when the WQN algorithm is applied on EEG signals perturbed by additive artifacts, it will transport the coefficients from the EEG + EOG/EMG into an EEG-like distribution, allowing restoration of the right statistics for the wavelet coefficients.

We conclude that the WQN algorithm can map artifacted signals to the smooth EEG distribution and thus remove possible jumps in the energy introduced by the artifacts, which are the cause of the monitoring issues we discuss below.


\section{WQN versus wavelet thresholding}\label{sec:wt_comparison}

\noindent We compared WQN algorithm with the wavelet thresholding methods that were initially introduced to denoise signals without smoothing sharp patterns. In this context, the noisy component is assumed to have small wavelet coefficients, sharing the same statistical properties at each scale.
The algorithm is efficient if the statistical properties of the signal to be recovered strongly differ from the noise e.g.\ if the wavelet coefficients of the signal form a {\sl sparse sequence}  (most of them almost vanish), and the nonzero coefficients are large. This situation is actually the opposite to the one considered in EEG artifact removal, where the artifacts to be eliminated have a sparse signature while the signal to be recovered  presents the homogeneous statistical properties of such a noise. However, wavelet thresholding and wavelet shrinkage also have been used in such contexts~\cite{krishnaveni2006removal,inuso2007waveletica,sweeney2012artifact,chavez2018surrogate} (but, once the splitting has been performed, the noisy part is kept instead of the sparse one); it is therefore enlightening to compare their performance with the WQN algorithm. One drawback of both of these algorithms is that they are {\sl local in the wavelet domain}, i.e.\ each wavelet coefficient is modified independently of the other ones, and therefore, they do not preserve the statistics of wavelet coefficients.

Non-local methods have been proposed (see e.g.\ block thresholding methods~\cite{Anton}), but they have the drawback of not restoring the correct coefficient statistics. This phenomenon has no negative impact when wavelet thresholding and wavelet shrinkage are used for their initial purpose, i.e.\ to restore the sparse part of the signal, but it becomes a major drawback when it is used for the opposite purpose of restoring the noisy component; for instance, on the specific position where the eliminated artifact was localized in time, the wavelet coefficients are set to 0, leading to inhomogeneities in the restored signal; one of the purposes of the WQN algorithm is to circumvent this drawback by restoring everywhere the correct anticipated statistics of the wavelet coefficients. As a consequence, it is not local in the wavelet domain: the value attributed to a coefficient depends on the whole statistic of coefficients at a given scale, and therefore the analysis of the regularity properties of the algorithm is more involved than for wavelet thresholding and wavelet shrinkage.

\begin{figure*}[t]
    \includegraphics[width=\textwidth]{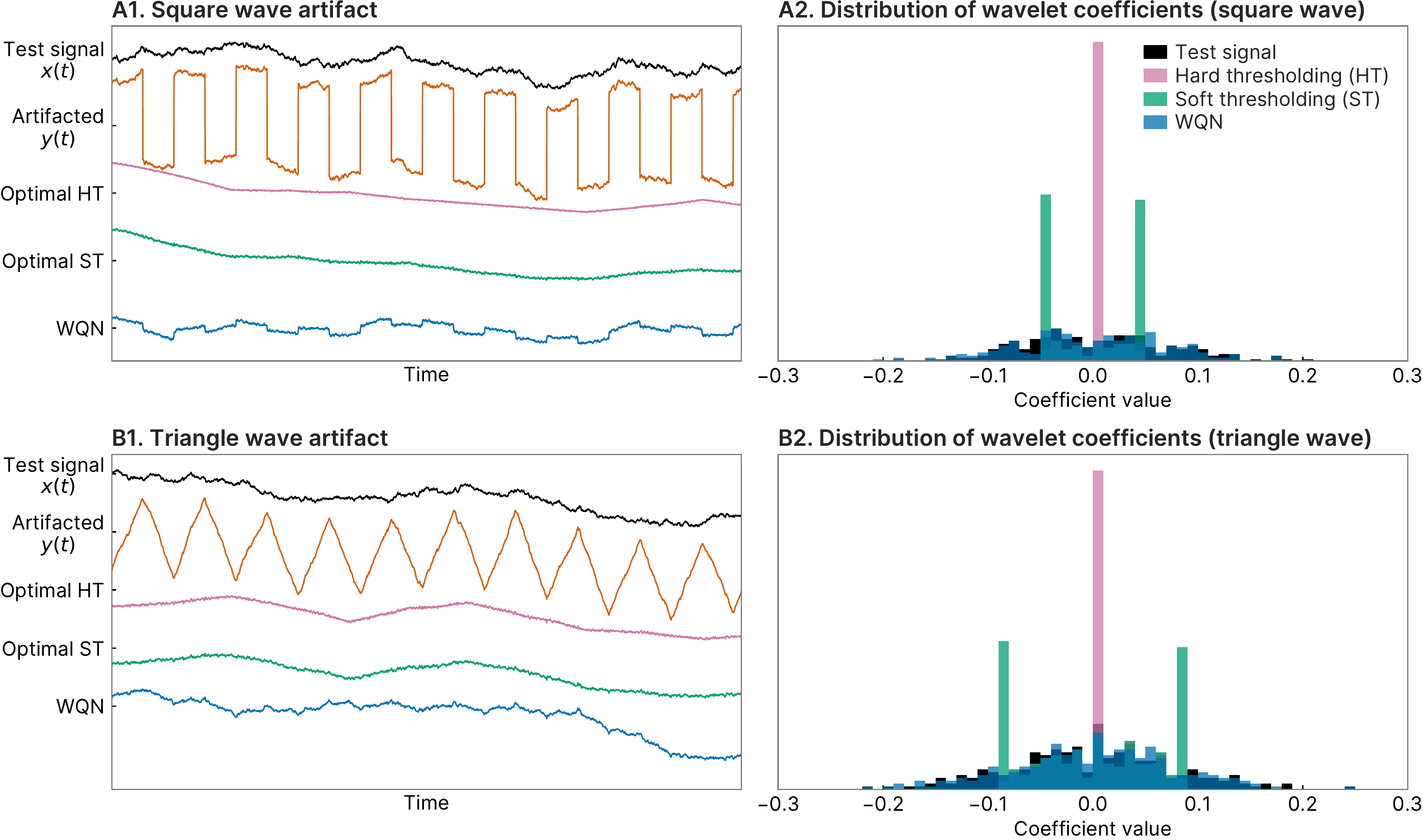}
    \caption{Comparison of wavelet coefficient statistics for hard thresholding (HT), soft thresholding (ST), and WQN artifact removal algorithms.
        \textbf{(A)}~Brownian motion signal corrupted by adding a square wave (A1) and distribution of the wavelet coefficients at level $m = 5$ (A2).
        \textbf{(B)}~Brownian motion signal corrupted by adding a triangle wave (B1) and distribution of the wavelet coefficients at level $m = 5$ (B2).
        The wavelet transport algorithm restores the correct wavelet coefficients statistics (A2, B2), in contrast with hard and soft thresholding which generate spurious peaks in the histogram respectively at the origin and at the threshold values.}%
    \label{fig:examplefbm}
\end{figure*}

At this stage, we carried out a series of simulations to compare the WQN algorithm with the wavelet thresholding methods.
To this aim, we considered a test signal $x(t)$ corrupted by an additive artifact $a(t)$, resulting in the artifacted signal $y(t) = x(t) + a(t)$.
To emulate an EEG signal we used a Brownian motion process for the test signal $x(t)$ (Fig.~\ref{fig:examplefbm}A). Note that the distribution of the wavelet coefficients of $x$ is a centered Gaussian of variance $2^{sm} $ and with short-range correlations (see~\cite{flandrin1992wavelet} for explicit values).
To simulate artifacts with different types of singularities, we chose $a(t)$ to be a square or a triangle wave (Figs.~\ref{fig:examplefbm}A1 and~\ref{fig:examplefbm}B1 respectively) with random phase shift to average out possible artifacts arising from the lack of translation invariance of the wavelet decomposition~\cite{coifman1995translation}.

We then applied wavelet thresholding and the WQN algorithms to the artifacted signal $y$ to recover a restored signal $\hat{x}$ (Fig.~\ref{fig:examplefbm}A). We consider hard and soft thresholding on the wavelet coefficients $w$, defined respectively by
\begin{align}\label{eq:threshold_hard}
     & \lambda^\text{hard}_\theta(w) = w \, \mathbbm{1}_{\left| w \right| < \theta}
    \\\label{eq:threshold_soft}
     & \lambda^\text{soft}_\theta(w) = \sgn(w)\, \min\left( |w|, \theta \right)
\end{align}
where $\theta$ is the threshold and $\mathbbm{1}$ is the indicator function. Note that these thresholding functions differ from the ones traditionally employed in wavelet denoising, as in the context of artifact removal one wants to suppress large coefficients.

To provide a comparison that does not dependent on the method used to define the threshold value $\theta$, we consider an ideal optimal thresholding algorithm minimizing the reconstruction error $\|\hat{x} - x\|_2^2$, i.e.\ for each scale $m$ we choose
\begin{equation}
    \theta^\text{opt}_m = \argmin_{\theta} \sum_n \| \lambda_\theta(c^y_{m,n}) - c^x_{m,n} \|_2^2
\end{equation}
where $c^y_m$ and $c^x_m$ denote the wavelet coefficients at scale $m$ of the artifacted ($y$) and original signal ($x$) respectively, and $\lambda_\theta$ is a thresholding function as defined in Eqs.~\ref{eq:threshold_hard}--\ref{eq:threshold_soft}. Note that this choice of the threshold cannot be employed in practice, as one does not know the clean signal $x$. However, we use this strategy in our simulations to provide the best reconstruction $\hat{x}$ achievable with wavelet thresholding in terms of $\ell^2$ distance from the signal $x$.

Additionally, we considered two commonly employed thresholding methods: universal thresholding \cite{johnstone1997wavelet} and SureShrink \cite{donoho1995adapting}. In universal thresholding, the threshold is defined as $\theta^\text{uni}_m = \sigma_m \sqrt{2 \log N}$, where $N$ is the signal length and $\sigma_m$ is the standard deviation of the wavelet coefficients of $x$ at scale $m$. Thresholds are chosen in a level-dependent way, one for each scale $m$. Coefficients are then shrinked by soft thresholding (Eq.~\ref{eq:threshold_hard}).
In SureShrink, the threshold value is chosen by minimizing the Stein's unbiased risk estimate (SURE) for the soft thresholding estimator of Eq.~\ref{eq:threshold_soft}. We implemented SureShrink with the hybrid scheme proposed in~\cite{donoho1995adapting}.

In Fig.~\ref{fig:examplefbm} we show an example of the application of the optimal thresholding and the WQN algorithms. Although in the WQN reconstruction some traces of the artifact are still visible, the original “texture” of the test signal $x(t)$ is better preserved compared to optimal thresholding which more aggressively suppresses higher frequencies (Fig.~\ref{fig:examplefbm}A1, B1).
In particular, WQN preserves well the distribution of the wavelet coefficients $c^x_{m,n}$, as opposed to hard thresholding (HT) which projects most of the large coefficients to zero, creating a peak at 0 in the distribution (Fig.~\ref{fig:examplefbm}A2,B2, pink), and soft thresholding (ST) which replaces them with the threshold value, creating two peaks at values $\pm \theta$ (Fig.~\ref{fig:examplefbm}A2,B2, green).
We conclude that HT and ST algorithms, with an optimal threshold in terms of $\ell^2$ distance, do not preserve the coefficient distribution, contrary to the WQN algorithm.

\begin{figure*}[t]
    \includegraphics[width=\textwidth]{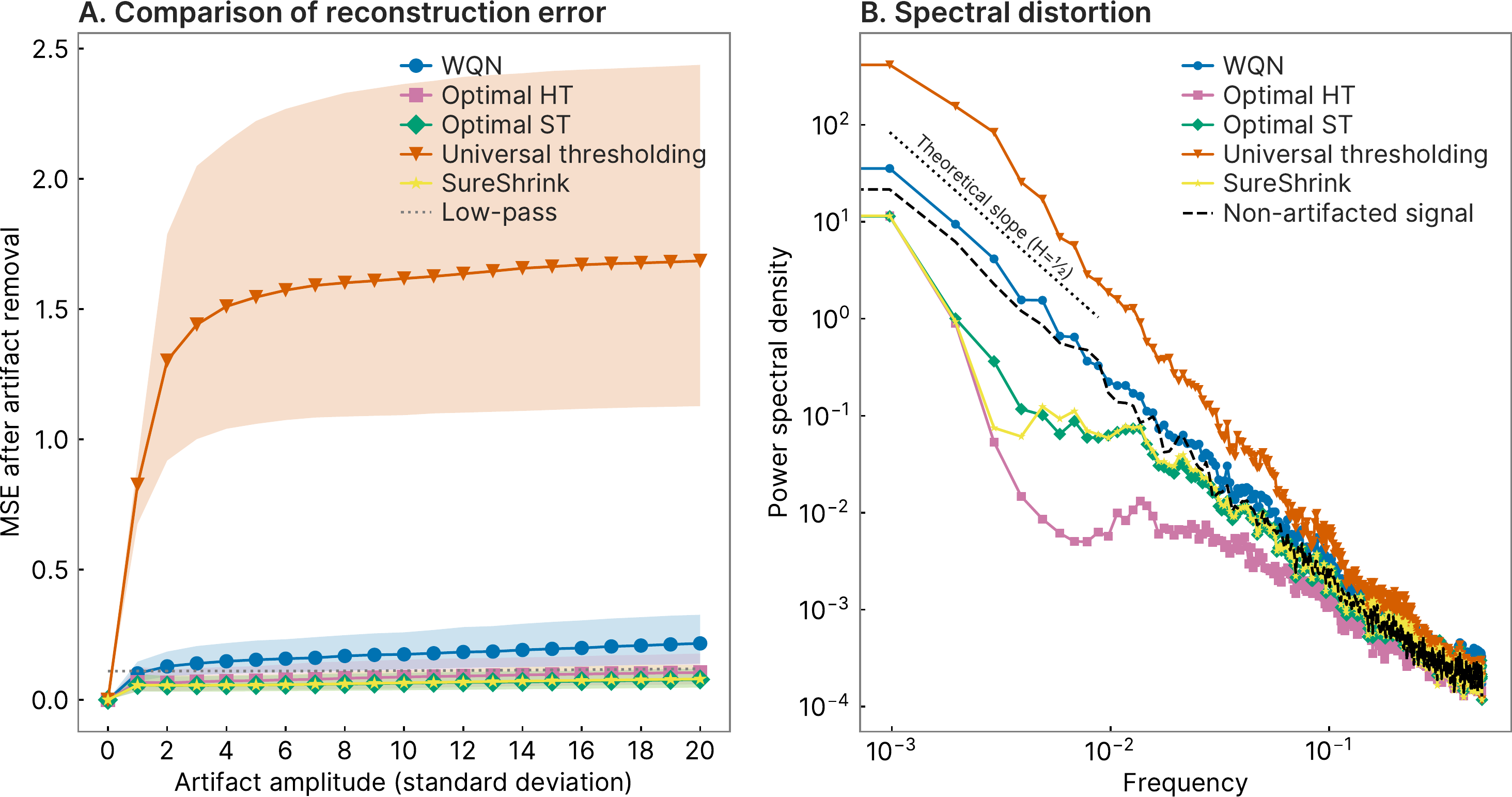}
    \caption{
        Comparison of reconstruction quality for wavelet thresholding algorithms and WQN.
        \textbf{(A)}~MSE after artifact removal for Brownian processes corrupted by additive square wave artifacts with varying amplitude. The lines represent the median value across \num{1000} realizations, with shaded areas indicating the interquartile range.
        \textbf{(B)}~Power spectral density of the signal restored by different methods. WQN restores the spectral scaling of the original Brownian signal ($\sim 1/f^2$), while thresholding algorithms, despite obtaining a lower MSE, do not properly recover the self-similarity of the signal.
    }%
    \label{fig:quantcomp}
\end{figure*}

In Fig.~\ref{fig:quantcomp}A we present the statistics of mean squared error (MSE) for square wave artifacts with varying amplitude. These results show that SureShrink is close to the optimal thresholding (Fig.~\ref{fig:quantcomp}A, yellow line), performing better than WQN (blue line) in terms of MSE, with the commonly used universal thresholding (orange line) showing the worse performance.
However, we note that the MSE minimization captures artifact removal performances only partially, as this could result in signal distortion due to aggressive suppression of the wavelet coefficients. As an extreme example, we report the MSE of a low-pass denoising algorithm which sets to zero all wavelet coefficients except the ones at the coarsest scale (Fig.~\ref{fig:quantcomp}A, gray dashed line). Such low-pass algorithm, while effective in terms of MSE minimization (especially for large artifacts), is clearly unsuitable for practical applications as it significantly distorts the signal.
To highlight such distortion, in Fig.~\ref{fig:quantcomp}B we show the spectral power density of the reconstructed signals $\hat{x}$ for the different methods. Optimal thresholding and SureShrink (pink, green, and yellow lines) significantly distort the power spectral density (PSD), breaking the self-similarity of the Brownian signal $x(t)$, while WQN preserves the scale relation, allowing for instance the estimation of the correct Hurst exponent ($H = 1/2$). The preservation of the self-similarity features is of particular importance in neurophysiological signals, which are often characterized based on the power-law trend ($1/f^\alpha$) in the PSD~\cite{donoghue2020parameterizing,wen2016separating}.

In Table~\ref{table:statistics} we report indicative metrics computed over multiple realizations of the Brownian signal $x(t)$ for the various algorithms.
These results highlight the different purposes of wavelet thresholding and the WQN algorithm. In the first case, one aims at reconstructing as closely as possible the original signal $x$ by minimizing the MSE $\mathbb{E} \left| \hat{x}(t) - x(t) \right|^2$. In the second case, one wishes to reconstruct the correct statistics of the signal $x(t)$ at each scale $m$, which we quantify by measuring the Wasserstein distance between the CDF of the wavelets coefficients $c^{\hat{x}}_m$, $c^x_m$ of the reconstructed and original signal $\hat{x}$, $x$ respectively, averaging on all decomposition scales.
For the universal thresholding and WQN algorithm, which retain the $1/f^\alpha$ scaling of the PSD (see Fig.~\ref{fig:quantcomp}B), we report the estimated Hurst exponents obtained by fitting the PSD.

\begin{table}
    \begin{tabularx}{\linewidth}{
            l|
            >{\raggedleft\arraybackslash}X
            >{\raggedleft\arraybackslash}X
            >{\raggedleft\arraybackslash}X
            >{\raggedleft\arraybackslash}X
            >{\raggedleft\arraybackslash}X
        }
        \toprule
                                              & \thead[r]{Ideal ST} & \thead[r]{Ideal HT} & \thead[r]{SureShrink} & \thead[r]{Universal WT} & \thead[r]{WQN}      \\
        \midrule
        \thead[l]{MSE (square)}               & \textbf{\num{0.07}} & \num{0.09}          & \num{0.07}            & \num{1.41}              & \num{0.15}          \\
        \thead[l]{MSE (triangle)}             & \textbf{\num{0.07}} & \num{0.09}          & \num{0.07}            & \num{1.19}              & \num{0.15}          \\
        \thead[l]{Avg Wasserstein (square)}   & \num{0.64}          & \num{0.84}          & \num{0.71}            & \num{2.07}              & \textbf{\num{0.26}} \\
        \thead[l]{Avg Wasserstein (triangle)} & \num{0.60}          & \num{0.80}          & \num{0.66}            & \num{1.79}              & \textbf{\num{0.26}} \\
        \thead[l]{Hurst exponent (square)}    & ---                 & ---                 & ---                   & \num{0.57}              & \textbf{\num{0.44}} \\
        \thead[l]{Hurst exponent (triangle)}  & ---                 & ---                 & ---                   & \num{0.69}              & \textbf{\num{0.48}} \\
        \bottomrule
    \end{tabularx}%
    \caption{Metrics of restoration for Brownian processes corrupted by additive square and triangle wave artifacts with standard deviation double than the original signal. For each metric we report the mean values over \num{1000} realizations. The Hurst exponent was recovered by fitting the power spectral density with $1/f^\alpha$, then using the relation $H = (\alpha - 1)/2$ (the Hurst exponent for Brownian motion is $H = 1/2$). Results for ST, HT, and SureShrink are not reported as the power spectral density could not be fitted by a $1/f^\alpha$ scaling (see for example Fig.~\ref{fig:quantcomp}). The best result for each metric is highlighted in black.}%
    \label{table:statistics}
\end{table}


\section{Regularity properties of the WQN algorithm}\label{sec:regularity}

\noindent A key property of wavelet bases is that, unlike the trigonometric system or other classical bases,  most classical functional spaces used in mathematical analysis, such as Sobolev spaces $H^{s,p}$ for $1< p < \infty$ or Besov spaces $B^{s, q}_p$ for $0< p, q \leq \infty$,  can be characterized by conditions on their coefficients~\cite{Mey90I}; this is referred to as the { \sl unconditional basis property} which plays a key role in statistics, see~\cite{donoho1995wavelet}. These spaces have a wavelet characterization which bears on the moduli of the wavelet coefficients, and which is an increasing function of each of these moduli. As an example, $f \in B^{s, q}_p$ if its wavelet coefficients satisfy the condition
\begin{equation}\label{eq:caracbes}
    \left(  \sum_n (2^{(1/p-1/2 -s)m} | c_{m,n}|)^p \right)^{1/p} \in l^q.
\end{equation}
Since, by construction,  the wavelet-transport algorithm is \emph{wavelet decreasing} (i.e.\ it does not increase the size of the wavelet coefficients, see Eq.~\ref{eq:coeff_normalization}), the following regularity property, also shared by the wavelet thresholding algorithms is satisfied: For  $0< p, q \leq  \infty$, and for any $s \in \RR$, its maps functions of a Besov or Sobolev space into the same space.  Note that this boundedness property also holds in the $S^\nu$ spaces, which were introduced in~\cite{Jaf9}, and are defined by conditions on the distributions of wavelet coefficients at each scale, and therefore supply a functional setting  particularly well fitted to the study of the WQN-wavelet transport algorithm. The same remark also applies to pointwise singularities, which can also be characterized by conditions bearing on the moduli of the wavelet coefficients, see e.g.~\cite{JaffToul}. To conclude, the regularity property above guarantees that the WQN algorithm does not introduce unwanted singularities, but on the contrary smoothens a signal contaminated by a sparse, highly non-Gaussian artifact.


\section{WQN to remove artifacts in real-time anesthesia monitoring}\label{sec:application}

\noindent During general anesthesia, real-time monitoring of the EEG signal provides continuous feedback about the depth of anesthesia, allowing to control the anesthetic dose required to keep the patient in a safe sleep state. Anesthesia monitoring machines are generally based on spectral features computed on time windows of 20--40 s~\cite{fahy2018technology}. When artifacts are detected, the data corresponding to whole windows are discarded and the output is suspended to prevent incorrect feedback, at the same time withdrawing fundamental information about the state of the patient. Being able to suppress transient artifacts in a continuous fashion, preserving the statistical properties of the signal, would thus allow a smoother monitoring.

\begin{figure*}
    \includegraphics[width=\linewidth]{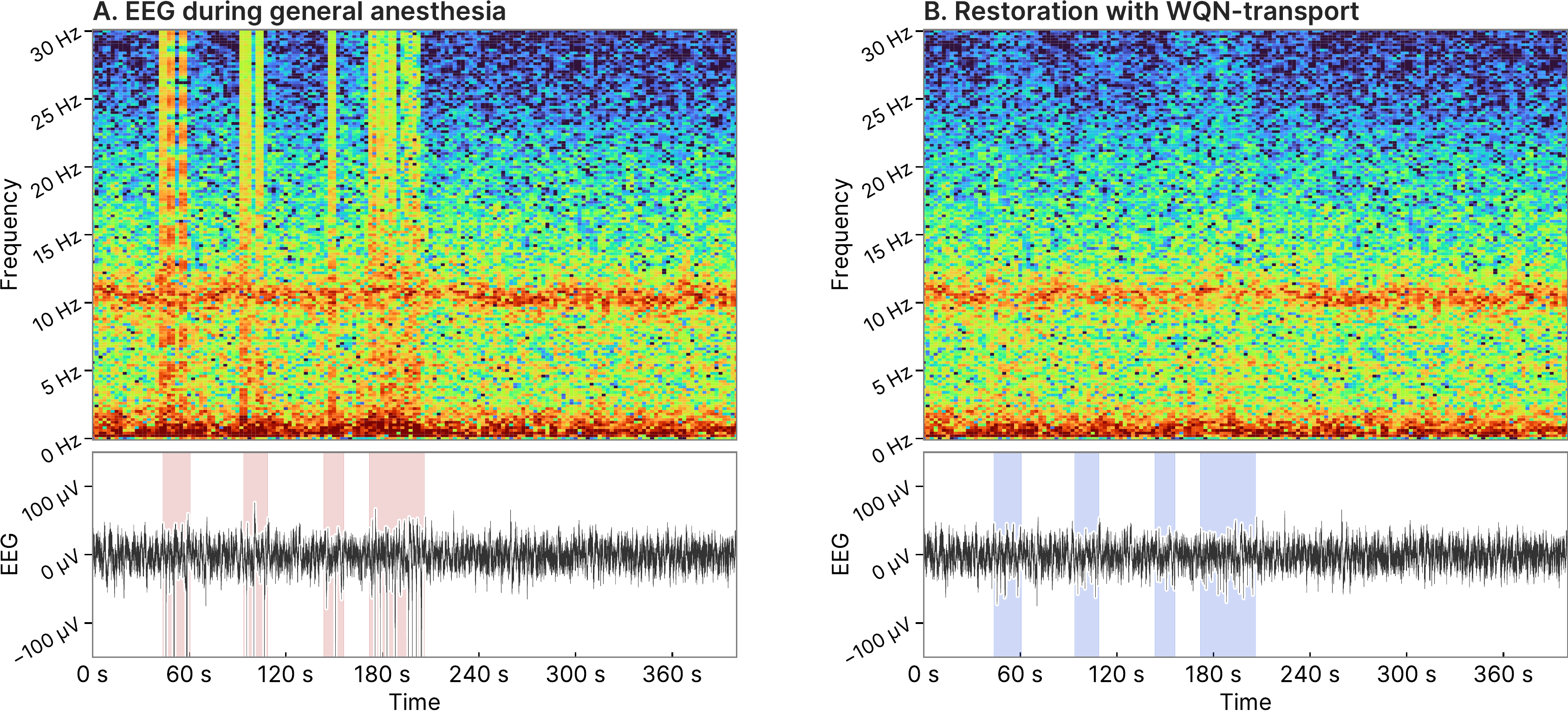}
    \caption{Example of restoring an EEG signal during general anesthesia.
        \textbf{A}~Spectrogram and EEG with artifact (shaded intervals). Artifacts are corrupting the spectrogram calculation and consequently the computations of parameters measuring the depth of anesthesia. \textbf{B}~Spectrogram and EEG after applying the WQN algorithm that has restored the artifacted regions. The spectrogram now clearly shows a continuous alpha band around 10 Hz.}%
    \label{fig:anesthesia}
\end{figure*}

While artifacts have typical timescale of seconds or less, they can affect computations on longer time intervals as spectral features are computed over relatively large sliding windows (20 to 40 seconds). To illustrate the performance of WQN algorithm, in Fig.~\ref{fig:anesthesia} we present an example of an EEG signal recorded during general anesthesia which is corrupted by electrode movement artifacts which consequently alter the computations of parameters measuring the depth of sedation. The artifacts are particularly visible on the spectrogram, with a large spread of the frequencies (Fig.~\ref{fig:anesthesia}A), compromising the quality of estimations. After applying the WQN algorithm, the artifacted regions present in the EEG and the spectrogram can be restored (Fig.~\ref{fig:anesthesia}B). Indeed, after WQN restoration, the spectrogram clearly shows a regular, continuous alpha band around 10 Hz.

\section{Conclusions}

\noindent The WQN algorithm allows to replace an artifacted epoch by a corrected signal, where the wavelet coefficient distribution is similar to the reference EEG.\@ In contrast with hard and soft thresholding methods, this algorithm does not require a threshold to be chosen. Moreover, WQN better preserves the signal self-similarity, allowing correct estimation of 1/f-like trends in the power spectrum, which are a well-known signature of brain signals.
In addition, WQN performs well on EEG signals recorded during anesthesia as there is usually an energy gap between the artifacts (characterized by large jumps and sudden fast oscillations) and the unperturbed EEG.\@ As shown in the present letter, the corrected signal is smoother than the artifact. This property is relevant because it allows direct automated computations from on the corrected signal that can be used to measure the depth of anesthesia. Clinical brain monitors are often programmed to stop operating following the detection of an artifact, as the artifact can undermine the computation of relevant parameters in the neighboring signal. We conclude that the WQN procedure could allow to improve the continuity of monitoring and real-time computation of the parameters associated to depth of sedation for general anesthesia.


\section*{Code availability}
\noindent The implementation of the algorithms described in this letter and the code used to generate all figures and results is made publicly available on Zenodo~\cite{dora2022wqn_code}.

\section*{Acknowledgements}
\noindent We thank the referee for several pertinent and helpful suggestions.


\bibliographystyle{elsarticle-num}
\bibliography{bibliography}

\end{document}